\documentclass[12pt]{iopart}

\newcommand{\st}{\mathrm{st}}
\usepackage{iopams}
\usepackage{graphicx}

\begin{document}

\title{On the emergence of large and complex memory effects in nonequilibrium fluids}

\author{A. Lasanta$^{1,4}$, F. Vega Reyes$^{2}$, A. Prados$^{3}$ and A. Santos$^{2}$}

\address{$^{1}$Gregorio Mill\'an Institute of Fluid Dynamics,
Nanoscience and Industrial Mathematics,
Department of Materials Science and Engineering and Chemical Engineering,
Universidad Carlos III de Madrid, 28911 Legan\'es, Spain}
\address{$^{2}$Departamento de F\'isica and Instituto de Computaci\'on Cient\'ifica Avanzada (ICCAEx), Universidad de Extremadura, 06006 Badajoz, Spain}
\address{$^{3}$F\'isica Te\'orica, Universidad de Sevilla, Apartado de Correos 1065, 41080
Sevilla, Spain}

\ead{$^{4}$alasanta@ing.uc3m.es}
\vspace{10pt}
\begin{indented}
\item Keywords:Complex fluids, Memory effect, Granular matter, Thermal
  behavior
\end{indented}

\begin{abstract}

  Control of cooling and heating processes is essential in many
  industrial and biological processes.  In fact, the time evolution of
  an observable quantity may differ according to the previous history
  of the system.  For example, a system that is being subject to
  cooling and then, at a given time $t_{w}$ for which the
  instantaneous temperature is $T(t_{w})=T_{\st}$, is suddenly put in
  contact with a temperature source at $T_{\st}$ may continue cooling
  down temporarily or, on the contrary, undergo a temperature rebound.
  According to current knowledge, there can be only one ``spurious''
  and small peak/low. However, our results prove that, under certain
  conditions, more than one extremum may appear. Specifically, we have
  observed regions with two extrema and a critical point with three
  extrema. We have also detected cases where extraordinarily large
  extrema are observed, as large as the order of magnitude of the
  stationary value of the variable of interest. We show this by
  studying the thermal evolution of a low density set of macroscopic
  particles that do not preserve kinetic energy upon collision, i.e.,
  a granular gas. We describe the mechanism that signals in this
  system the emergence of these complex and large memory effects, and
  explain why similar observations can be expected in a variety of
  systems.
\end{abstract}

\section{Introduction}

Experimental observations reveal that the response to an
excitation of complex condensed matter systems may depend on the
entire system's history, and not just on the instantaneous value of
the state variables
\cite{KAHR79,RF03,PM06,X10,FFS14,HCCW17,SCKL17,nagel_experimental_2017}. This is usually
called \textit{memory effect}. Memory effects signal the breakdown of
the thermodynamic (or hydrodynamic or macroscopic, depending on the
physical context) description.  Some typical memory effects include
shape memory in polymers \cite{X10}, aging and rejuvenation in spin
glasses \cite{B02}, active matter \cite{JKL17}, and polymers
\cite{S80}, and the counterintuitive Mpemba effect \cite{MO69,LVPS17,JANUS18}.

One of the most relevant memory effects related to thermal processes
was originally observed by Kovacs and collaborators \cite{KAHR79} in a
polymer system, which was subject to quenching to a low temperature
$T_{1}$ from an equilibrium state at temperature $T_{0}>T_1$. After a
long enough waiting time $t_{w}$, but still relaxing towards
equilibrium at $T_{1}$, the temperature was suddenly increased back to
an intermediate value $T_{\st}$, $T_{1}< T_{\st} < T_{0}$, such that the
instantaneous value of the volume $\mathcal{V}(t=t_w)$ equalled the
equilibrium value $\mathcal{V}_{\st}$ corresponding to $T_{\st}$.
Subsequently, the volume $\mathcal{V}(t)$ did not remain flat but
followed a nonmonotonic evolution. This nonmonotonic behavior,
denominated later as \textit{Kovacs hump}, consists in reaching
\textit{one} maximum before returning to its equilibrium value
$\mathcal{V}_{\st}$.

We have described above the typical \textit{cooling} procedure, but
also a \textit{heating} protocol can be considered
($T_{0}<T_{\st}<T_{1}$), for which $\mathcal{V}(t)$ exhibits a single
minimum at $t>t_w$. Quite recently, Kovacs-like memory effects have
been thoroughly investigated in glassy systems
\cite{MS04,prados_kovacs_2010}, granular fluids
\cite{PT14,plata_kovacs-like_2017}, active matter \cite{KRI17}, and
disordered mechanical systems \cite{lahini_nonmonotonic_2017}. The
memory effect is typically quite small: the maximum deviation of
$\mathcal{V}(t)$ from the stationary value $\mathcal{V}_{\st}$ is
several orders of magnitude smaller than $\mathcal{V}_{\st}$
\cite{KAHR79,PT14,plata_kovacs-like_2017,MS04,KRI17}.

One of the main aims of our work is to show that the actual memory
effects landscape is in general far more complex than expected. First,
we show that several extrema---instead of only one---may appear in a
single heating/cooling protocol {\em \`a la} Kovacs, contrary to what
has been previously observed
\cite{MS04,prados_kovacs_2010,PT14,plata_kovacs-like_2017,KRI17,lahini_nonmonotonic_2017}. Second,
very large memory humps, of the order of magnitude of the stationary
value of the quantity of interest, can be observed. To the best of
our knowledge, both features have not yet been reported in the
literature. It must be noted that humps much larger than those
predicted by linear response theory have recently been found in a
nonlinear active matter model~\cite{KRI17}, but the relative
deviation from the steady state is still of a few hundredths therein.

Our results are found in a granular fluid but the mechanism presented
for these features is quite general. Thus, giant and complex memory
effects---not necessarily of the Kovacs-type---may be expected to appear
in many natural and artificial systems. These memory effects have
obviously important implications in problems like, for instance, system
stabilization.

\section{Description of the system and theoretical solution}

We consider a collection of identical solid spheres at low particle
density so that collisions are always instantaneous and binary but
inelastic, i.e., energy is not conserved and we deal with a granular
gas \cite{H83,AT06}. In this case, particles have homogeneous mass
density and we employ the rough hard sphere collisional model with
constant coefficients of normal and tangential restitution, $\alpha$
and $\beta$, respectively, which is quite realistic for a variety of
materials at low particle density \cite{FLCA94}.

Let us discuss first why the granular gas of rough spheres is a good
candidate for eventually finding complex memory effects. Memory
effects appear always in complex systems that consist of many
structural units, for which a continuum description seems in principle
appropriate. Within this kind of description, the instantaneous value
of the complete set of macroscopic variables completely characterizes
the system's time evolution \cite{B72}. However, there are states that
cannot be completely described only with the system macroscopic
variables, and it is precisely for these states where a memory effect
can be observed. As a matter of fact, this kind of distinct states for
which the macroscopic description fails are theoretically very well
understood in the context of the kinetic theory of gases \cite{B72}.

Furthermore, the granular gas of rough spheres can have extremely long
relaxation times before it falls into a state where the macroscopic
description is valid  \cite{VSK14,VS15}, giving room to the emergence
of eventual long lasting memory effects. And, most importantly, in
this kind of system there are always two intrinsic, independent, and
potentially large temperature scales---the translational and rotational
granular temperatures---with a highly nonlinear coupling. All these
facts open new spaces in the search of novel important features in complex memory effects, including eventually multiple extrema.

To keep things simple, we consider the granular gas to be in a
spatially homogeneous state at all times. The translational velocities
are denoted by $\mathbf{v}$, while the angular (or rotational)
velocities are denoted by ${\bomega}$.  The system is thermalized by
a stochastic but homogeneous volume force \cite{WM96,OLDLD05}
characterized by a noise intensity $\chi_0^2$ (see \ref{app_A}).

The kinetic description of our system starts from the corresponding
Boltzmann--Fokker--Planck equation for the granular gas under this kind
of forcing \cite{VS15} (see \ref{app_A}). The exact solution to this
kinetic equation can be formally expressed by means of an expansion
around a Maxwellian distribution with variances $T_t$ (translational
temperature) and $T_r$ (rotational temperature) in the translational
and angular velocities, respectively. The total granular temperature
is given by $T=(T_t+T_r)/2$, which is proportional to the mean
kinetic (translational plus rotational) energy per particle. By adopting a
dimensionless time scale $\tau$, proportional to the number of
collisions per particle (see \ref{app_A}), the evolution equations for the
temperatures can be written as

\begin{equation}\label{evol_gamma}
\frac{\partial \ln\theta(\tau)}{\partial \tau}=\frac{2}{3}\left[\mu_{20}(\tau)-\mu_{02}(\tau)-\gamma(\tau)\right],
\end{equation}
\begin{equation} \label{evol_gamma2}
\frac{\partial\ln\gamma(\tau)}{\partial \tau}=\mu_{20}(\tau)-\gamma(\tau).
\end{equation}
Above, $\theta\equiv T_r/T_t$ is the temperature ratio and
\begin{equation}
  \gamma \equiv \left(\frac{T_{\mathrm{noise}}}{T}\frac{1+\theta}{2}\right)^{\frac{3}{2}}
\end{equation}
is a dimensionless measure of the noise intensity, where
$T_{\mathrm{noise}}\equiv
m(3\chi_0^2/4\sqrt{\pi}n\sigma^2)^{\frac{2}{3}}$ with $n$ being the
particle density, and $m$ and $\sigma$ being the mass and diameter of
a sphere, respectively.  The reduced collisional moments $\mu_{20}$
and $\mu_{02}$ (see \ref{app_A} for more reference) are functionals of
the whole velocity distribution and therefore the above system of
equations is not closed. In order to solve it, we use the
\textit{first Sonine approximation}, which refers to the first
nontrivial truncation of the aforementioned exact infinite expansion
\cite{VSK14}. For this, together with Eqs.\ (\ref{evol_gamma}) and
(\ref{evol_gamma2}), we need to incorporate the evolution equations
for the fourth-order cumulants and the initial values of $\gamma$,
$\theta$, and these cumulants (see \ref{app_A}).

We generate a common initial state for all the temperature evolution
curves we subsequently analyze. At an arbitrary time, which we choose
to be the time origin $\tau=0$, and over an arbitrary previous
microscopic state, we apply an instantaneous thermal pulse to the
granular gas. In this way, the rotational modes ($T_r$) of the
granular gas are quenched, whereas the translational modes ($T_t$) are
subject to a large heating. As a result, most of the initial kinetic
energy is in the translational modes, so that the total initial
temperature is $T(0)=T_t(0)/2$ and the temperature ratio is
$\theta(0)=0$. Moreover, all the fourth-order cumulants vanish because
the initial distribution that results from the heat pulse is a
bi-variate ($T_r, T_t$) Maxwellian. By this procedure, the system
forgets all the previous thermal history of the system, assuring
always the same nonequilibrium initial state.

\begin{figure}[t]
  \centering
  \includegraphics[width=0.6 \columnwidth]{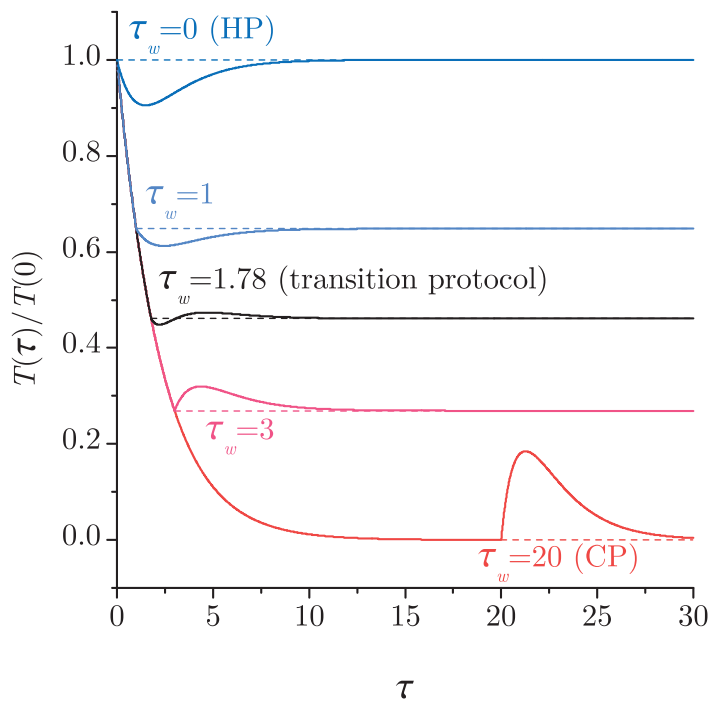}
  \caption{{Illustration of the protocols considered in this work.}
    The granular gas is prepared in an initial state ($\tau=0$) for
    which all the energy is concentrated in the translational degrees
    of freedom, as described in the text. In a first stage,
    $0<\tau<\tau_{w}$, the granular gas freely cools. Then, at the
    waiting time $\tau=\tau_{w}$, the noise intensity is suddenly
    increased from zero to a value such that the instantaneous
    temperature $T(\tau_{w})$ coincides with the corresponding steady
    temperature $T_{\st}$. The curves shown for $\tau>\tau_w$
    correspond to the so-called \textit{normal} Kovacs response.  Time
    $\tau$ measures the average number of collisions per particle.
    Note that, in order to visualize the Kovacs effect, the relative
    deviations of $T(\tau)$ from $T_{\st}$ in the response curves have
    been magnified by a factor $r=5$ for all the protocols, except for
    the transition one, for which $r=100$. All the curves correspond
    to normal and tangential restitution coefficients $\alpha=0.8$ and
    $\beta=0$, respectively.}
 \label{protocols_fig}
 \end{figure}

From the initial state we have just characterized, the granular gas is
left to cool freely, due to the intrinsically inelastic particle
collisions \cite{H83}, for a waiting time $\tau_w$. At
$\tau=\tau_{w}$, we suddenly apply the stochastic force, with an
intensity such that the corresponding steady temperature $T_{\st}$ to
be reached equals the instantaneous temperature value at the moment of
turning the noise on, i.e., $T_{\st} = T(\tau_w)$. If $T(\tau>\tau_w)$
further departs from $T_{\st}$, then a Kovacs-like \textit{memory
  effect} is observed. What we call \textit{protocol} is the thermal
procedure that we have just described. Depending on the waiting time
$\tau_w$ for turning the stochastic heating on, the system spans
different classes of temperature evolution curves. This is depicted
and explained in Fig.~\ref{protocols_fig}. For the sake of simplicity,
we investigate the two limiting cases in Fig.~\ref{protocols_fig};
i.e., $\tau_w=0$ (heating protocol, HP) and $\tau_w\to\infty$ (cooling
protocol, CP).

In the CP, since the system is left cooling down for a long time, the
system is already in the \textit{homogeneous cooling state} (HCS)
\cite{H83} at $\tau_w$. In the HCS, the temperature $T(t)$ is the only
relevant variable and decays in time following Haff's law \cite{H83},
whereas the temperature ratio and the fourth-order cumulants are time
independent, and their values depend only on the parameters $\alpha$
and $\beta$ \cite{VSK14}. Therefore, the conditions for this protocol
at $\tau_w$ are
$\gamma(\tau_w)=\gamma_{\st}[(1+\theta_{\st})/(1+\theta_{\mathrm{HCS}})]^{\frac{3}{2}}$,
$\theta(\tau_w)=\theta_{\mathrm{HCS}}$, and the fourth-order cumulants
at $\tau=\tau_w$ also equal their HCS values.

In the HP, the initial conditions for the Kovacs experiment are
different. Since we turn on the stochastic force right after the
thermal pulse, the initial conditions are those of a bi-variate
Maxwellian. Therefore, we have that
$\gamma(\tau_w)=\gamma_{\st}(1+\theta_{\st})^{\frac{3}{2}}$,
$\theta(\tau_w)=0$, and, in addition, all the cumulants vanish at
$\tau=\tau_w$.

\section{Results and Discussion}

\begin{figure}[t]
  \centering
  \includegraphics[width=0.6\columnwidth]{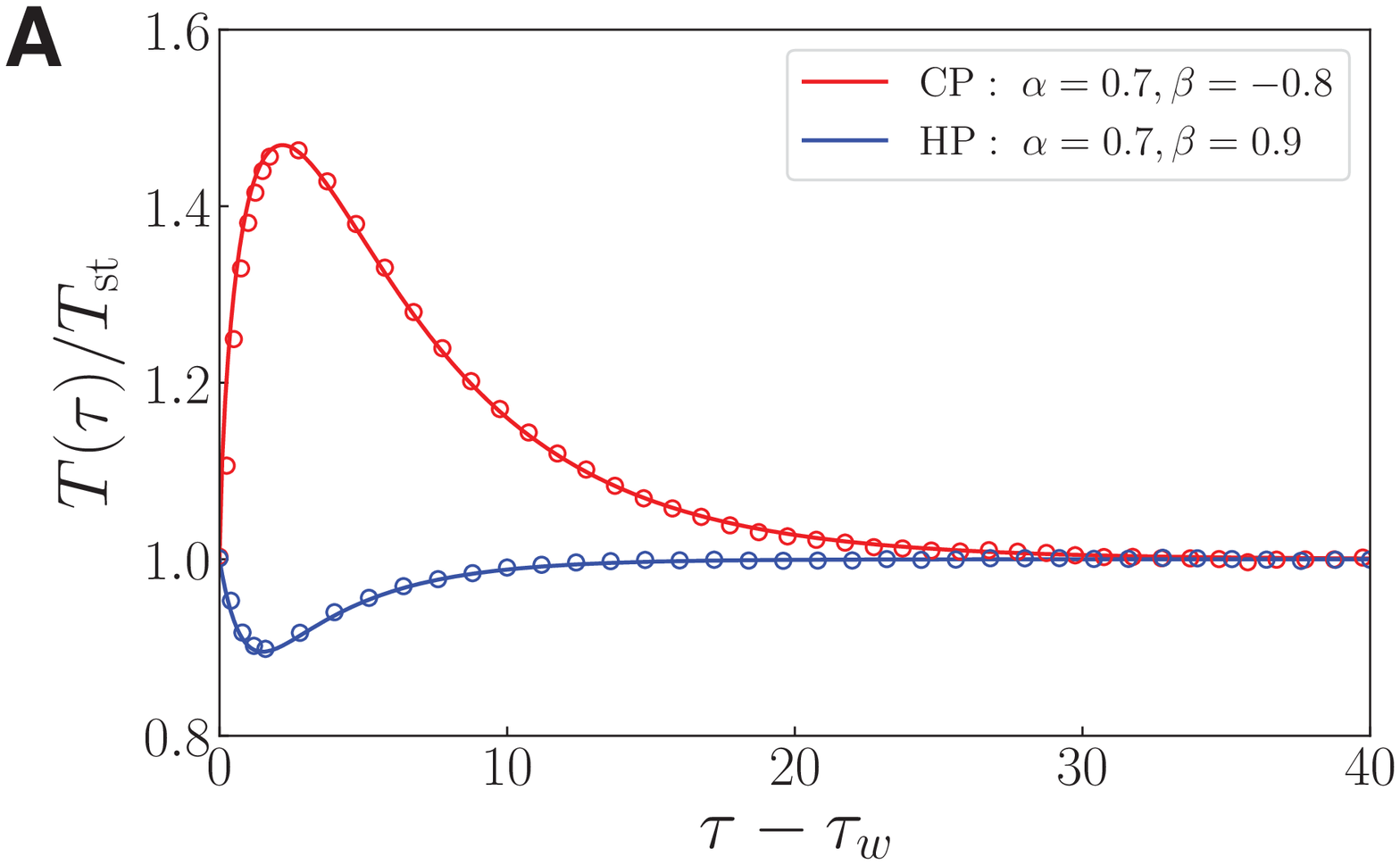}
  \includegraphics[width=0.6\columnwidth]{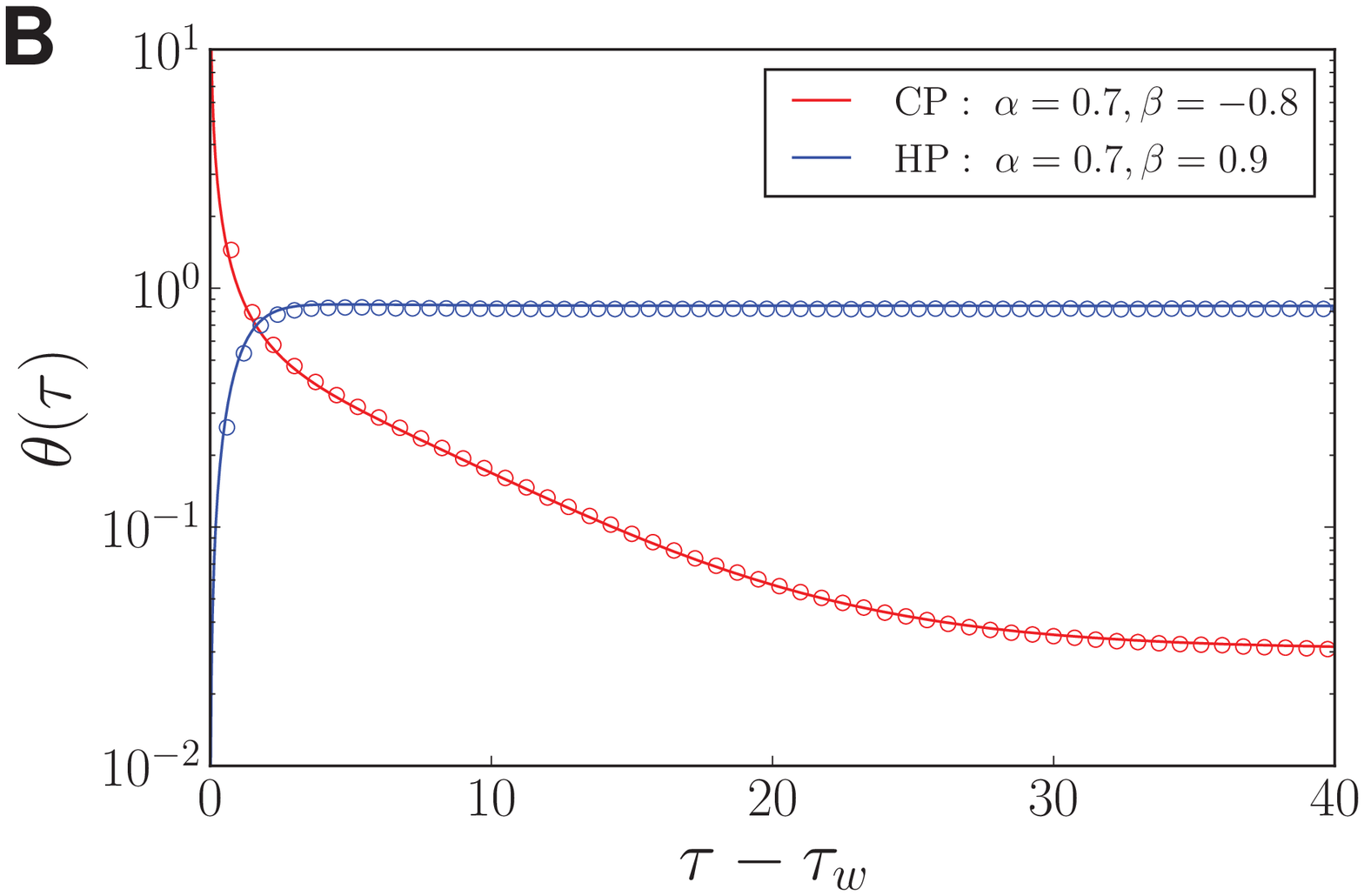}
  \caption{{Large Kovacs humps in the granular gas.} Panel \emph{A} shows two
    examples of macroscopic Kovacs humps for a granular gas with
    $\alpha=0.7$. The upper curve corresponds to the CP (with
    $\beta = 0.9$) whereas the lower curve corresponds to the HP (with
    $\beta = -0.8$), as measured in MD simulations. Panel \emph{B} shows the time evolution of the corresponding rotational-to-translational temperature ratio $\theta=T_r/T_t$. The simulation
    results show an almost perfect agreement with our theoretical
    predictions (lines).\label{colossal}}
 \end{figure}

 Two data sets from molecular dynamics (MD) simulations (see
 \ref{app_B}) of the granular gas for both the HP and the CP, together
 with their corresponding theoretical predictions, are represented in
 Fig.~\ref{colossal}\emph{A}, which clearly shows the appearance of
 very large memory effects. The temperature humps displayed here, of
 approximately $100\%$ for the CP and $10\%$ for the HP, are larger by
 at least two orders of magnitude than previously observed memory
 effects in athermal systems, which at most range from a few
 thousandths to a few hundredths of the stationary value of the
 relevant variable \cite{PT14,KRI17}.  The theoretical curves
 displayed in Fig.~\ref{colossal}\emph{A} have been obtained by means
 of a bi-variate \textit{Maxwellian approximation}, in which all the
 cumulants are assumed to be zero (see \ref{app_A}). Thus, the
 essential property driving the giant memory effect here is the
 existence of two independent temperature scales, translational and
 rotational, i.e., the breakdown of equipartition as given by the fact
 that $\theta\neq 1$. This is further illustrated in
 Fig.~\ref{colossal}\emph{B}, which shows $\theta(\tau)$ for the same
 cases as in Fig.~\ref{colossal}\emph{A}. Again, the agreement between
 theory and simulation is excellent, even at the level of the two
 contributions to the total temperature. Note that the relaxation time
 in the CP case is much longer than in the HP one.

\begin{figure*}[!ht]
  \centering
\includegraphics[width=0.6\columnwidth]{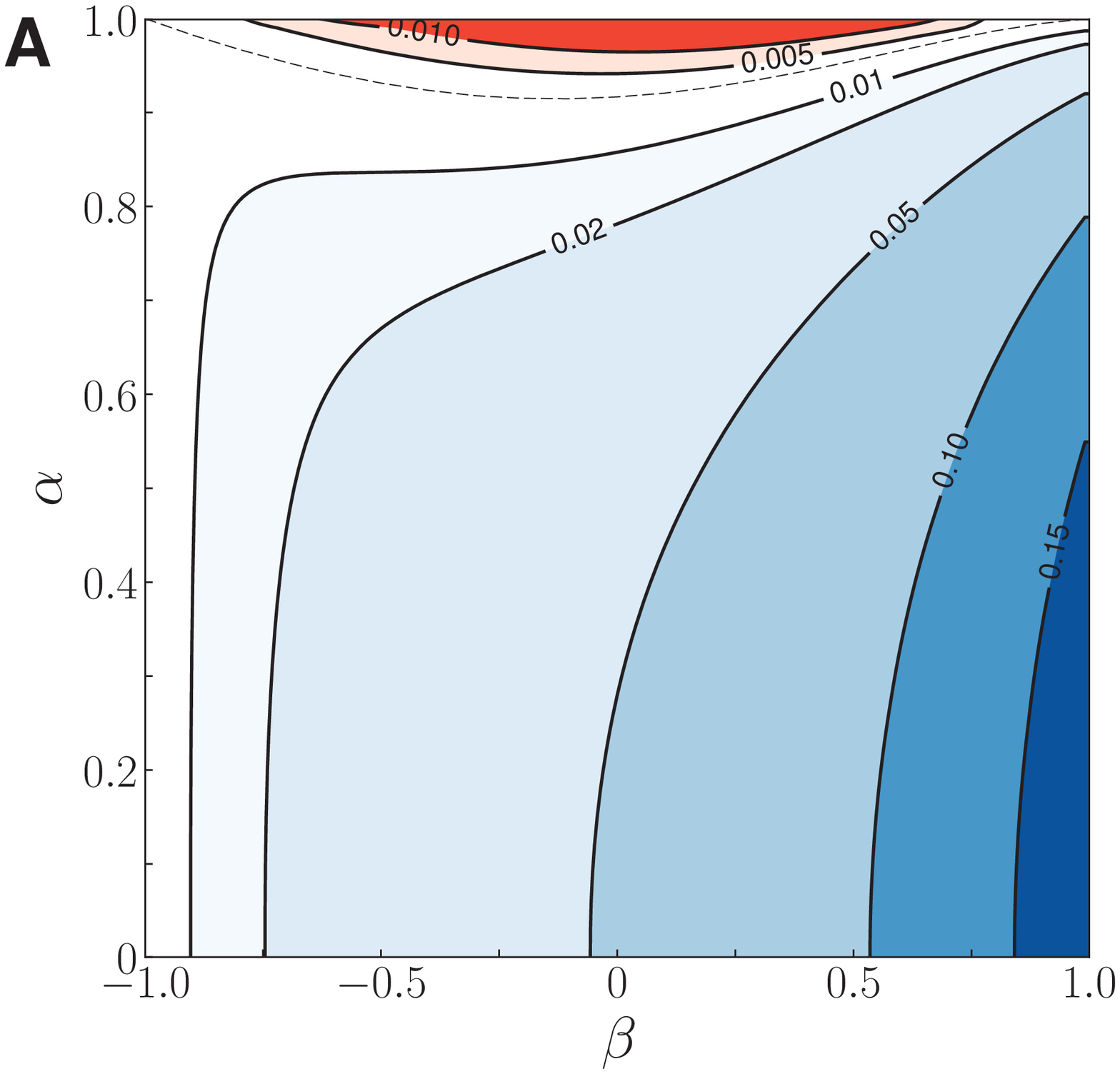}\quad\quad  \includegraphics[width=0.6\columnwidth]{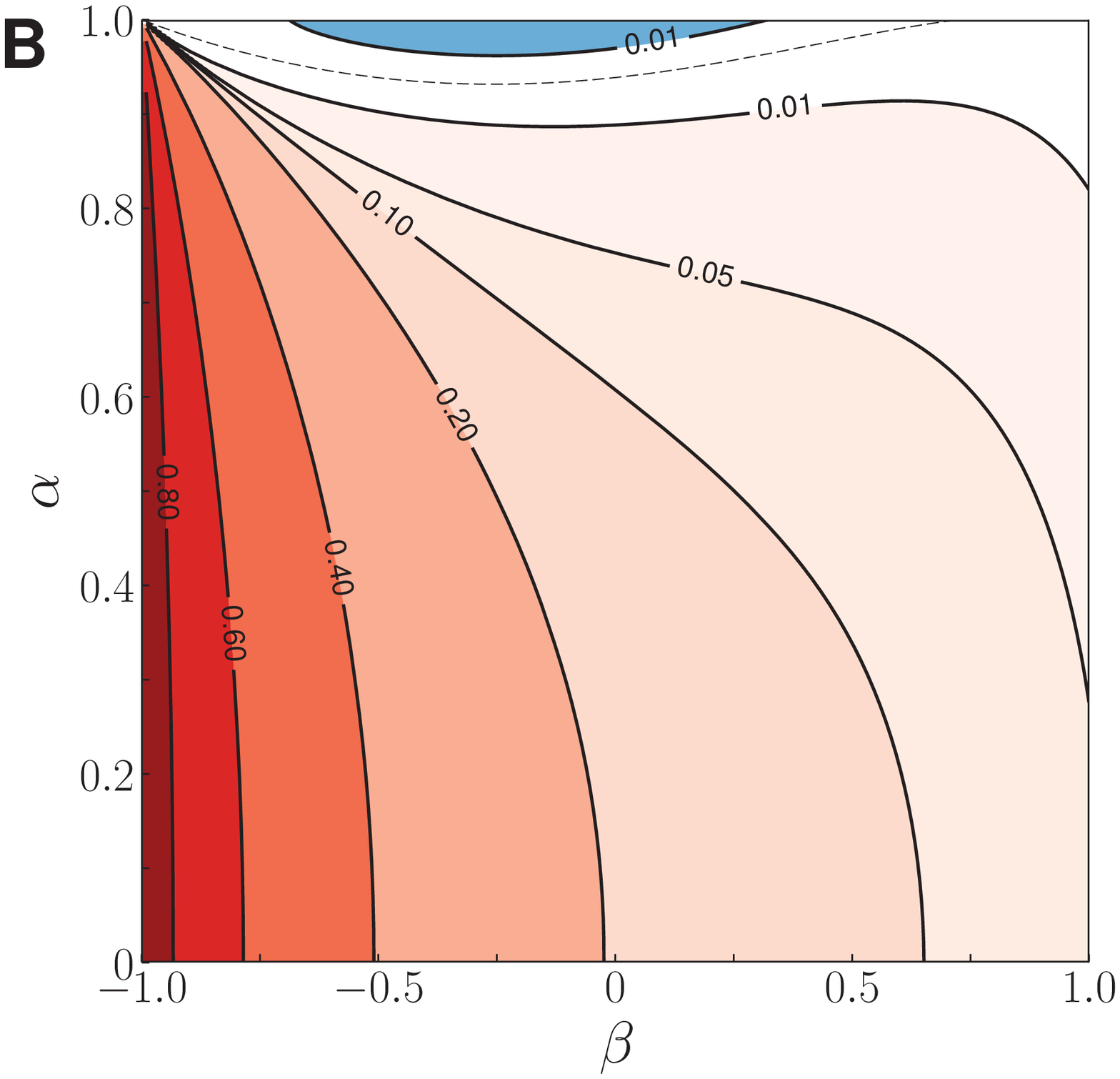}
\caption{{ Contour plots of large extrema in the Kovacs response.}
  Large minima ($|\mathcal{H}_{m}|$) are represented by bluish
  contours and large maxima ($\mathcal{H}_{M}$) by reddish
  contours. Dashed lines indicate the $\mathcal{H}_M=|\mathcal{H}_m|$
  transition curves, for which the predominant extremum changes sign,
  from maximum to minimum and vice versa. Above and below these curves
  we find $\mathcal{H}_M> |\mathcal{H}_m|$
  ($\mathcal{H}_M< |\mathcal{H}_m|$) and $\mathcal{H}_M<\mathcal{H}_m$
  ($\mathcal{H}_M>|\mathcal{H}_m|$) behaviors, respectively, in the HP
  (CP). (\emph{A}) Heating protocol (HP). (\emph{B}) Cooling protocol
  (CP).
    \label{HMm}
  }
\end{figure*}

 Let us denote the earliest minimum and maximum in the temperature
 evolution as $T_{m}$ and $T_{M}$, respectively. We also define
 $\mathcal{H}_{m}\equiv T_{m}/T_{\st}-1<0$,
 $\mathcal{H}_{M}\equiv T_{M}/T_{\st}-1>0$, accordingly. In
 Fig.~\ref{HMm} we present contour plots highlighting the regions with
 large $|\mathcal{H}_m|$ (HP normal response, CP anomalous response)
 and $\mathcal{H}_M$ (HP anomalous response, CP normal response). We
 also plot the transition line $\mathcal{H}_M=|\mathcal{H}_m|$ from
 normal to anomalous response. Huge Kovacs humps appear, especially in
 the normal region for the CP, in which the size of reported humps can
 be as large as $100\%$, relative to the steady temperature.

 In order to characterize and quantify complexity in the thermal
 response we define the parameter $\mathcal{S}$,
\begin{equation}
\label{complexity}
  \mathcal{S} =\mathrm{sgn} ({\mathcal{H}_{1}})
  \frac{\min(|\mathcal{H}_{m}|,\mathcal{H}_{M})}
  {\max(|\mathcal{H}_{m}|,\mathcal{H}_{M})},
\end{equation}
where $\mathcal{H}_{1}$ (equal to either $\mathcal{H}_{m}$ or
$\mathcal{H}_{M}$) is the magnitude of the earliest extremum.  Note
that $\mathcal{S}=0$ if there is only one extremum. Thus,
$\mathcal{S}\neq 0$ is the signature of the emergence of more complex
response, i.e., with more than one extremum, in the normal-to-anomalous
transition. In the transition region, $|\mathcal{S}|$ attains its
maximum value, $|\mathcal{S}|=1$, when both extrema are of the same
size and neither dominates. The sign of $\mathcal{S}\in [-1,1]$ is
equal to that of the earliest extremum, providing further information
on the detailed structure of the response.

\begin{figure*}[t]
  \centering
  \includegraphics[width=0.48\columnwidth]{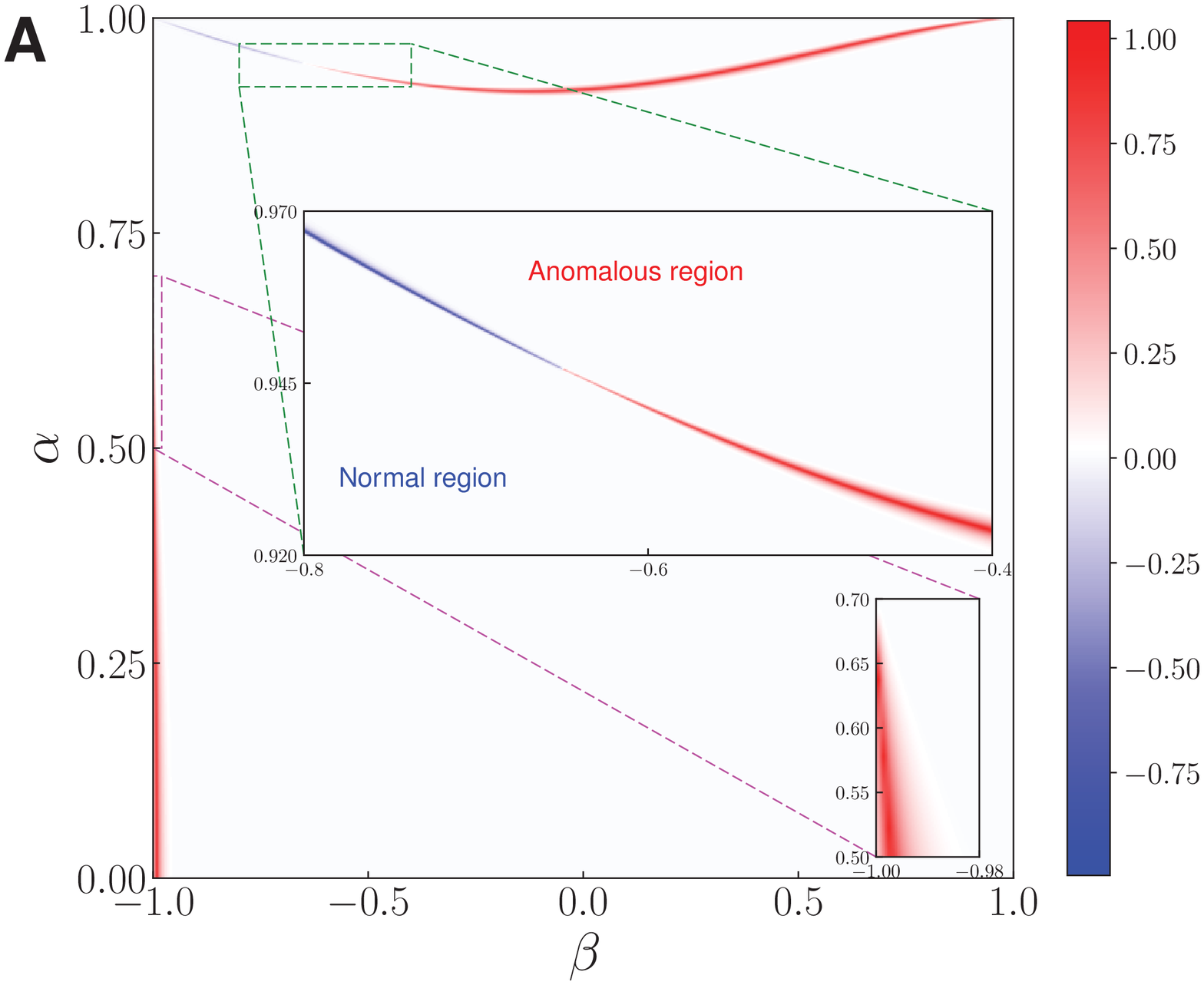}
  \includegraphics[width=0.48\columnwidth]{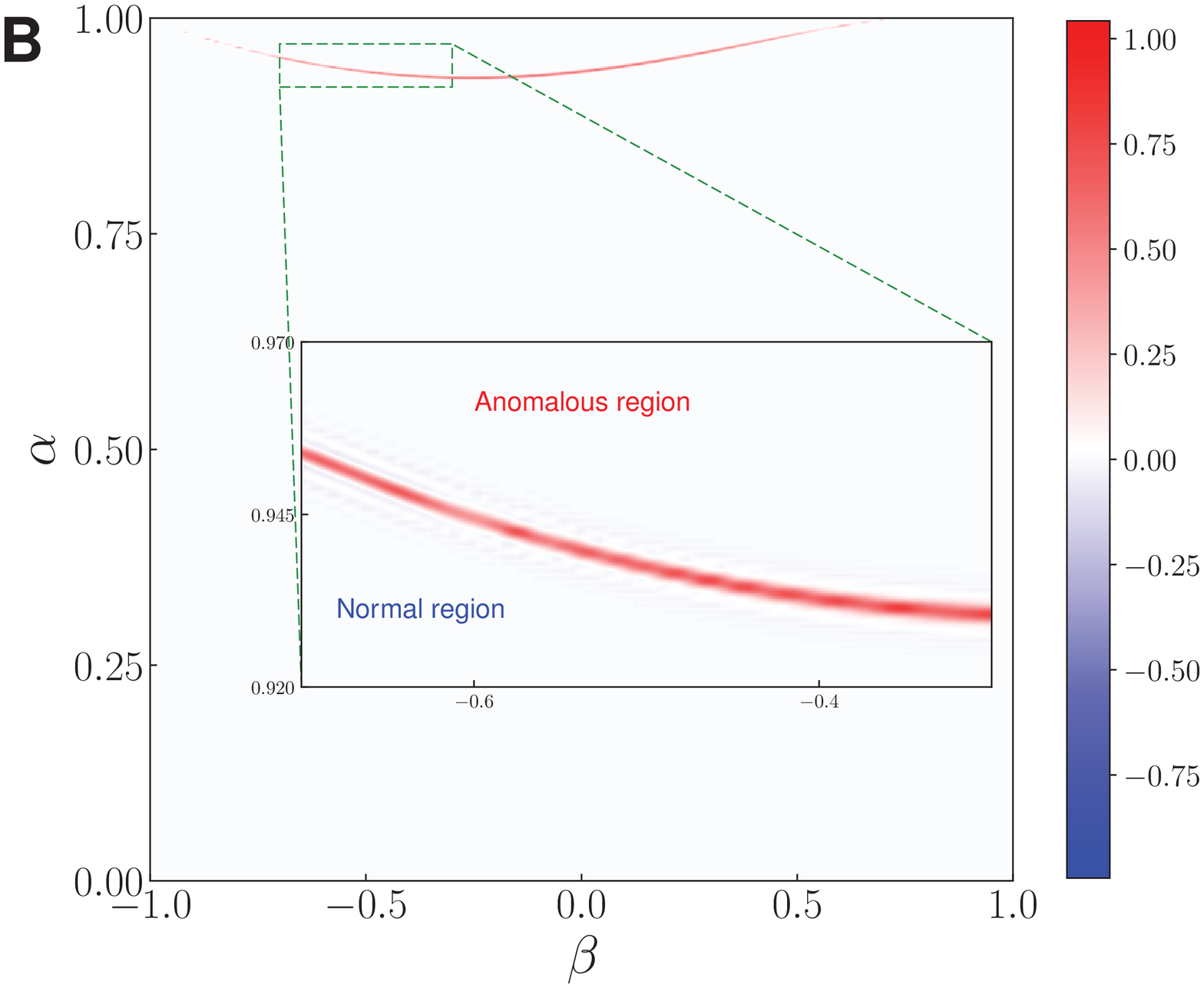}
  \begin{minipage}{0.4\textwidth}
     \includegraphics[width=0.7\columnwidth]{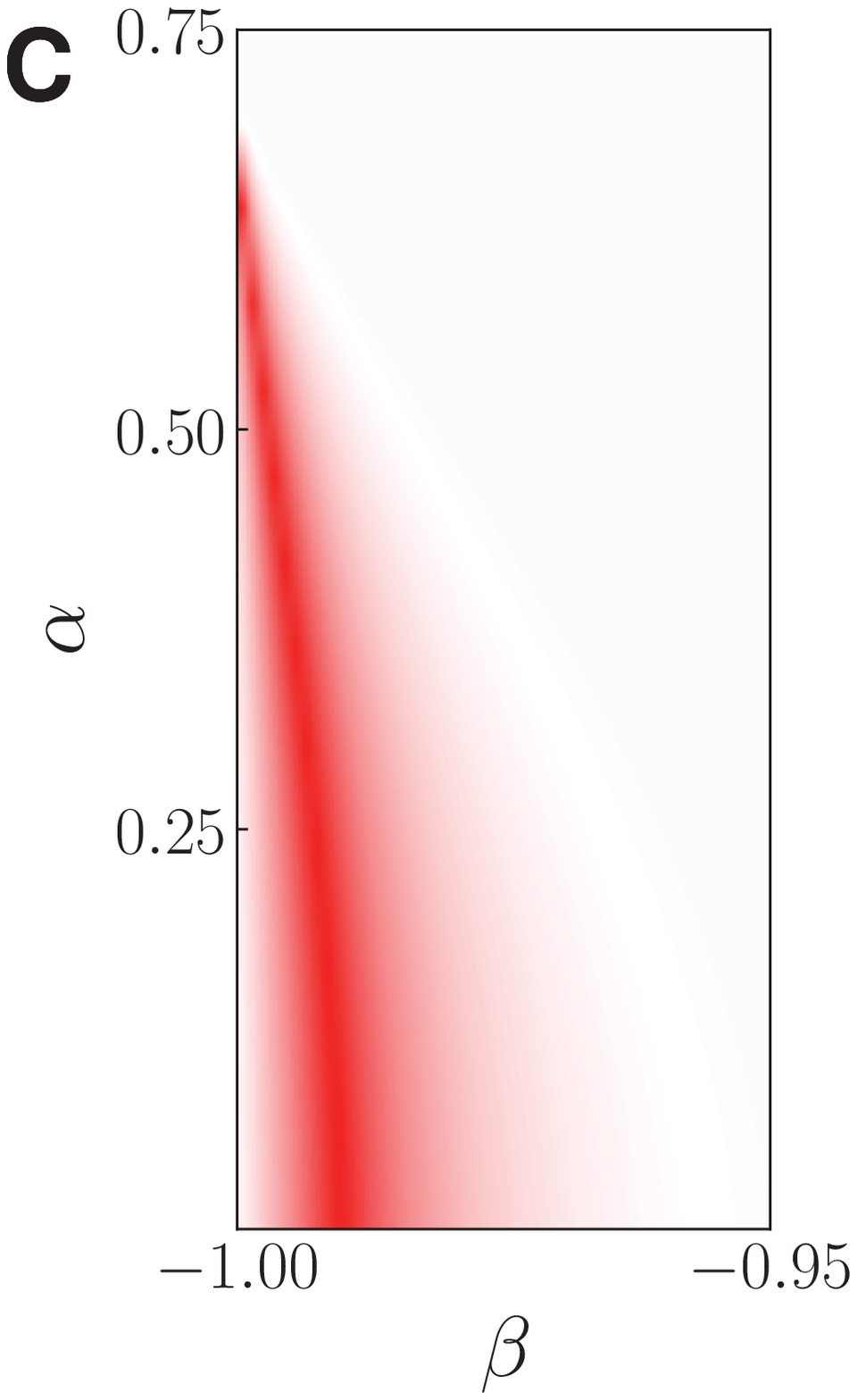}
  \end{minipage}
  \begin{minipage}{0.58\textwidth}
\includegraphics[width=0.8\columnwidth]{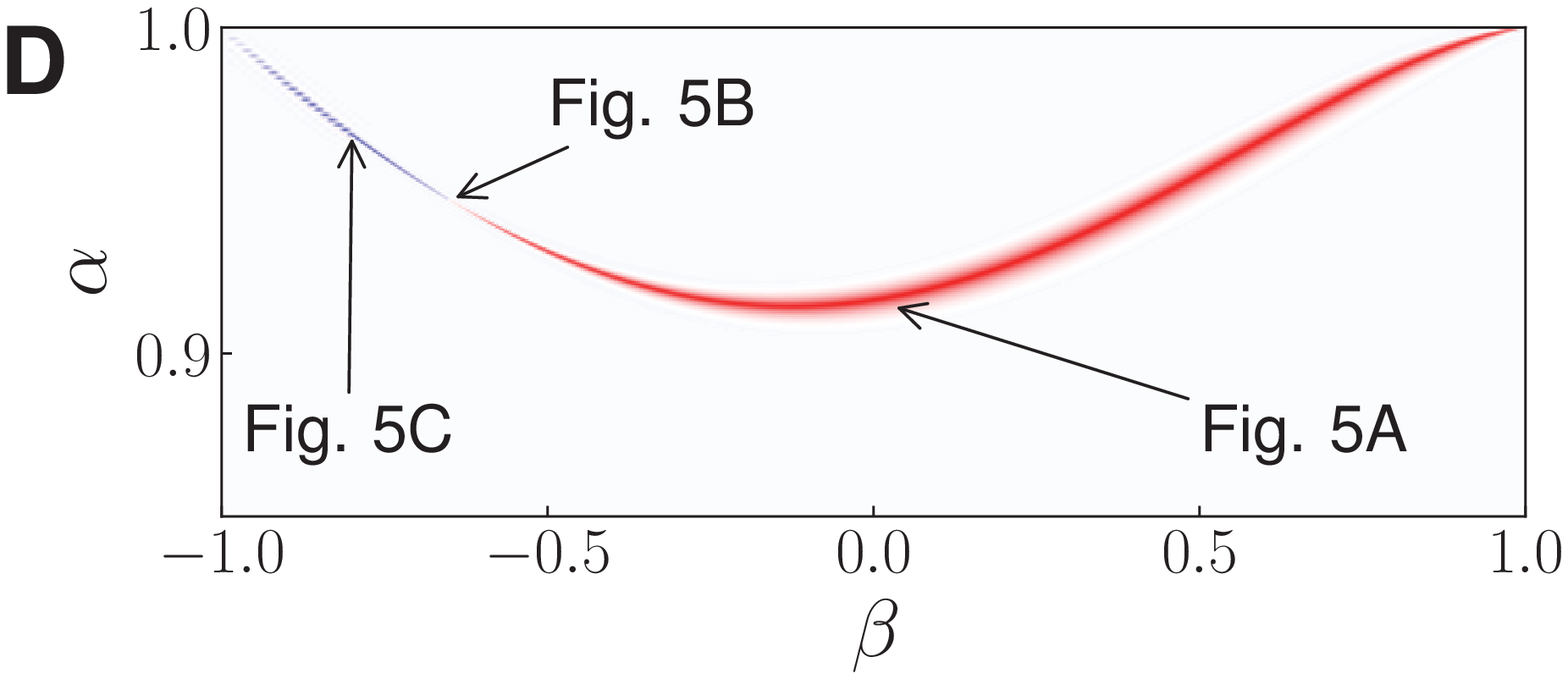}
\includegraphics[width=0.8\columnwidth]{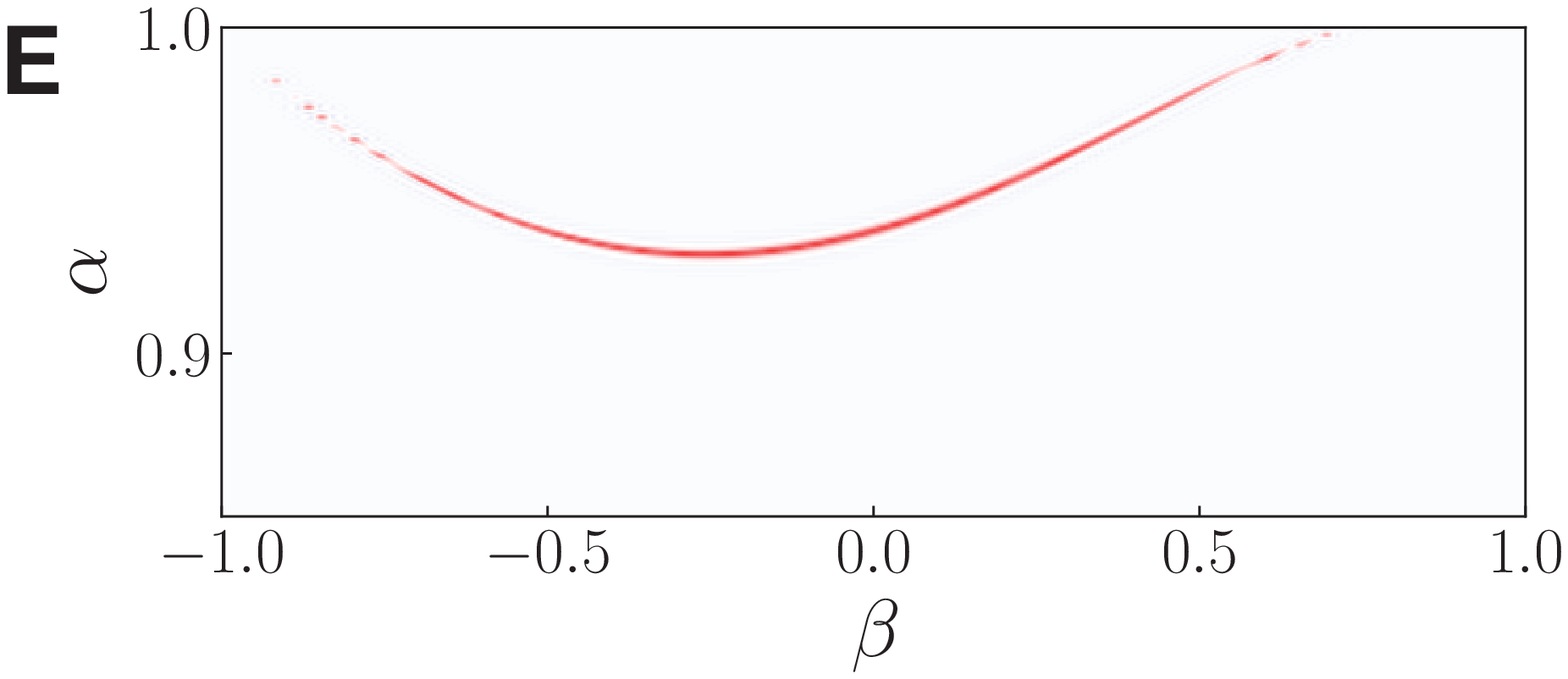}
  \end{minipage}
  \caption{{Kovacs complexity ($\mathcal{S}$) phase diagrams.}
    Density plot of $\mathcal{S}$ vs. the $\alpha, \beta$ complete
    space parameter: (\emph{A}) for the HP; (\emph{B}) for the CP.
    (\emph{C, D}) Insets of complex memory regions next to the smooth
    and the quasielastic limits, respectively, for the HP. (\emph{E})
    Inset of the complex memory region next to the quasielastic limit
    for the CP. In the HP, as seen in panels \emph{A} and \emph{D},
    three different types of transition exist: $\mathcal{S}<0$
    (bluish, $\beta<-0.65$), $\mathcal{S}>0$ (reddish,
    $\beta> -0.65$), and the intermediate mechanism
    ($\mathcal{S}\approx 0$) for $\beta\approx-0.65$, as depicted
    below in Fig.~\ref{transitions}\emph{B}. However, in the CP, see
    panels \emph{B} and \emph{E}, only a $\mathcal{S}>0$-type
    transition has been observed.
    \label{skewness}
  }
\end{figure*}

Figure~\ref{skewness} represents $\mathcal{S}$ as a function of the
coefficients of restitution $\alpha, \beta$, by solving the system of
Eqs.\ (\ref{evol_gamma}) and (\ref{evol_gamma2}). Panels \emph{A},
\emph{C}, and \emph{D} correspond to the HP, whereas panels \emph{B}
and \emph{E} correspond to the CP. We have highlighted in blue (red)
regions with $\mathcal{S}<0$ ($\mathcal{S}>0$), whereas all points
with ``simple'' memory behavior, i.e., $\mathcal{S}=0$, remain
white. The complex regions are thin but still occupy noticeable
sections of the parameter space, especially taking into account that
they fall into ranges of experimental values of $\alpha$ and $\beta$
commonly present in a variety of materials \cite{FLCA94}. In
Fig.~\ref{skewness}\emph{A} (HP), we clearly observe two zones rich in
complex memory effects. In panel \emph{C}, the first complex zone is
zoomed in. This region is attached to the smooth limit,
$\beta\sim -1$, and only displays $\mathcal{S}>0$ for high
inelasticities, up to $\alpha=1/\sqrt{2}$. In panel \emph{D}, the
second complex region is zoomed in. Within this region, which is close
to the quasielastic limit $\alpha\sim 1$, the system displays both
$\mathcal{S}> 0$ and $\mathcal{S}< 0$ behavior.  In
Fig.~\ref{skewness}\emph{B} (CP), only one complex Kovacs region next
to the quasielastic limit, inside which $\mathcal{S}> 0$, has been
identified. Panel \emph{E} shows a close-up thereof.

\begin{figure}[!ht]
  \centering
  \includegraphics[width=0.6\columnwidth]{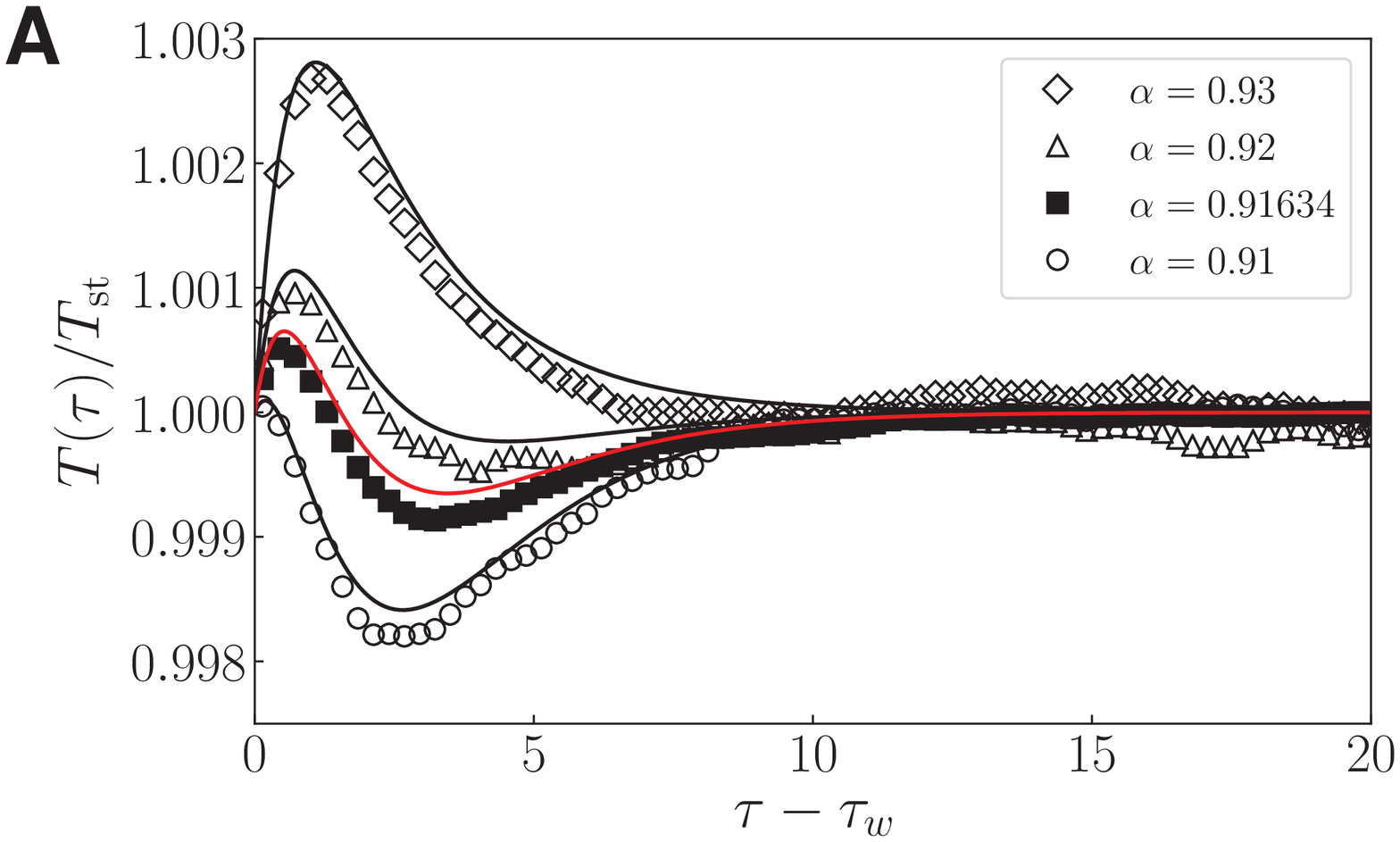}
  \includegraphics[width=0.6\columnwidth]{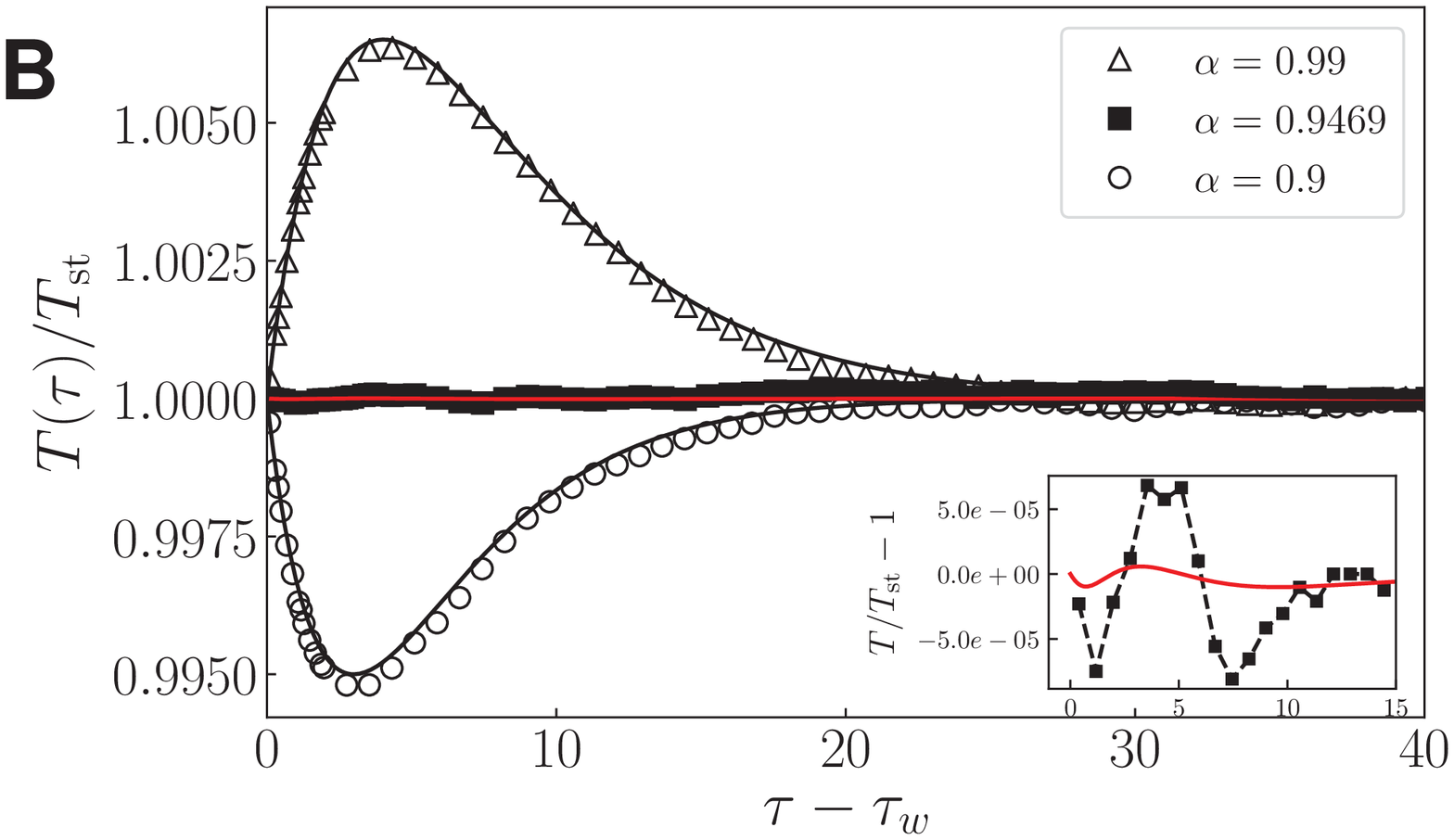}
  \includegraphics[width=0.6\columnwidth]{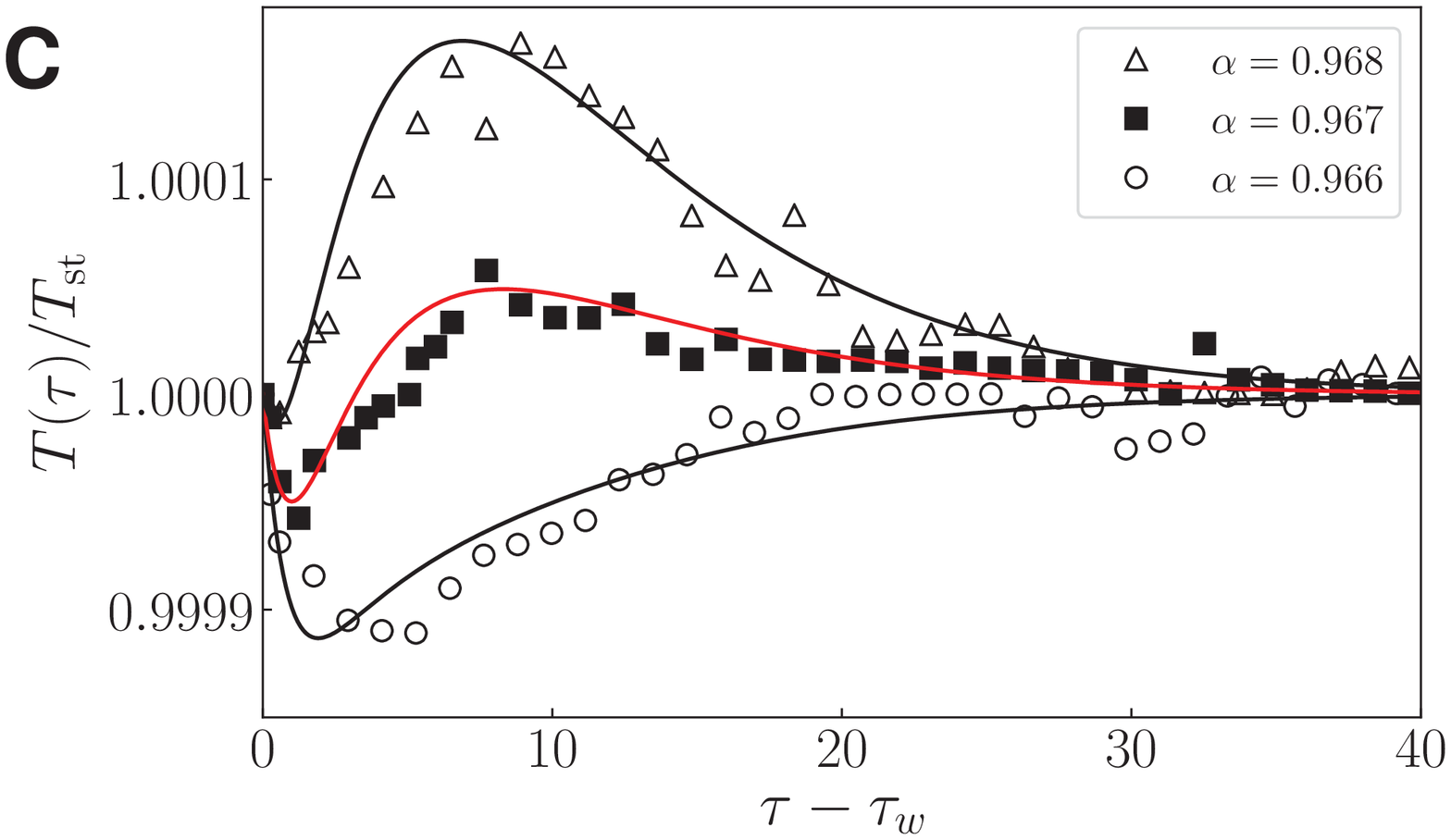}
  \caption{{Mechanisms for the transition from normal to anomalous in
      the HP.} There are three of these mechanisms, which are shown
    here by DSMC simulations (symbols) and our theoretical approach
    (lines). Specifically, through (\emph{A}) $\mathcal{S}>0$ at
    $\beta = 0$, (\emph{B}) $\mathcal{S} \approx 0$ (actually a triple
    Kovacs hump transition mechanism; see inset, where the line
    joining the simulation points is a guide to the eye) at
    $\beta = -0.65$, and (\emph{C}) $\mathcal{S} < 0$ at
    $\beta = -0.8$.  In order to assist in locating these transitions
    in the parameter plane $(\alpha,\beta)$, their positions have been
    annotated in Fig.~\ref{skewness}\emph{D}.
     \label{transitions}}
 \end{figure}

It is important to mention that we have found that all the details of
the complex regions emerge in the theoretical solution only when the
cumulants are taken into account. This indicates that the temperatures
$T_t$ and $T_r$ do not explain in full detail by themselves the
complexity of memory effects found in the rough granular gas. Let us
also point out that we have found for the HP a critical narrow region
with a discontinuous transition from $\mathcal{S}> 0$ to
$\mathcal{S}<0$, which is signaled in panel \emph{D} in
Fig.~\ref{skewness} and represented in time evolution curves in
Fig.~\ref{transitions}.

In this critical region, the system displays several different
mechanisms for the transition from complex to simple---only one
extremum---behavior. The latter can be either the normal behavior of
molecular systems \cite{KAHR79,B02,MS04} (also present in
nonequilibrium systems) or the anomalous behavior exclusive of
nonequilibrium systems \cite{PT14,KRI17}. This is appropriately tagged
in panels {\emph{A}} and {\emph{B}} of Fig.~\ref{skewness}, in which
we have labeled the corresponding normal and anomalous regions. In the
narrow critical region, $\mathcal{S}$ discontinuously jumps from
(small) negative to positive values and three consecutive temperature
extrema appear before stabilization in the stationary value is
attained. Otherwise, $\mathcal{S}$ has a well-defined sign and the
transition from complex to simple is continuous.

 Figure~\ref{transitions} displays the evolution curves of the
 temperature for the three different Kovacs transitions that we have
 found, in all cases depicted here for the HP: the $\mathcal{S}>0$
 transition in panel \emph{A}, the $\mathcal{S} \approx 0$ transition
 in panel \emph{B} (in its inset we show the three consecutive humps),
 and finally the $\mathcal{S} < 0$ case in panel \emph{C}. All
 theoretical curves are compared against the numerical solution of the
 kinetic equation, obtained by means of the direct simulation Monte
 Carlo (DSMC) method (see \ref{app_B}). The agreement is in general
 excellent, which once more shows the accuracy of our theoretical
 approach. Although the size of the humps in the transition regions
 appear smaller than those in Fig.~\ref{colossal}\emph{A} with simple
 memory behavior, yet they are of the same order of magnitude as those
 previously reported in the smooth granular gas \cite{PT14}.

\section{Conclusions}

Our work puts forward a general mechanism for the emergence of
significantly large memory effects. \textit{Enormous humps} can be
expected if the time evolution of the system under scrutiny is
controlled by at least two independent and comparable in magnitude
physical variables (here the translational and rotational
temperatures) but with only one (here the total temperature) being
relevant for the macroscopic or hydrodynamic description. In addition,
\textit{complex Kovacs response}, with more than one extremum, can be
expected if the time evolution of the system depends on several
additional relevant variables. Here, these additional variables are
the fourth-order cumulants, whose sometimes nonmonotonic
relaxation~\cite{VSK14} probably enhances memory effect complexity.

So far, and despite the large number of previous works on analogous
phenomena, only one extremum in the Kovacs response has been
reported. In thermal systems, in which the usual
fluctuation-dissipation theorem holds and the stationary (equilibrium)
distribution has the canonical shape, this is consistent with linear
response results that predict normal behavior with only one
maximum~\cite{prados_kovacs_2010}. In athermal systems, the Kovacs
response also includes anomalous behavior, but once more only one
extremum has been observed~\cite{PT14,KRI17}. Therefore, an interesting
prospect is elucidating whether or not the nonlinear theoretical
framework developed in Ref.~\cite{KRI17} allows for complex response
with more than one extremum.

Memory effects of the size and complexity we have observed here can
potentially be present in other athermal or molecular systems. Several
variables of comparable magnitude must be coupled in their time
evolution in nonlinear form, even if only a subset thereof is relevant
in the macroscopic description.  This may be relevant, for instance,
in active matter systems, where nonlinear effects are important in
general \cite{KRI17,MJR13}. We think our results are also especially
significant for future experimental work, since we expect these large
memory effects to be measurable in granular dynamics experiments; a
thermally homogeneous system may be achieved by means of homogeneous
turbulent air fluidization \cite{OLDLD05}.

\subsection*{Acknowledgements} The authors thank Prof.\ J.\ S.\ Urbach for fruitful discussions. This
  work has been supported by the Spanish Agencia Estatal de
  Investigaci\'on Grants (partially financed by the ERDF)
  No.~MTM2017-84446-C2-2-R, ~MTM2014-56948-C2-2-P (A.L.), and
  No.~FIS2016-76359-P (F.V.R. and A.S.), and also by Universidad de
  Sevilla's VI Plan Propio de Investigaci\'on Grant PP2018/494
  (A.P.). Use of computing facilities from Extremadura Research
  Centre for Advanced Technologies (CETA-CIEMAT), funded by the ERDF
  is also acknowledged.

\appendix

\section{Theory}
\label{app_A}

The stochastic force ($\mathbf{F}^{\mathrm{wn}}$) has the form of a white noise: $\langle {\bf
  F}_i^{\mathrm{wn}} (t) \rangle = {\bf 0}$, $\langle {\bf F}_i ^{\mathrm{wn}} (t) {\bf
  F}_j ^{\mathrm{wn}} (t') \rangle = \mathsf{I}\,m^2 \chi_0^2
\delta_{ij}\delta(t-t')$, where indexes $i,j$ refer to particles,
$\mathsf{I}$ is the $3\times 3$ unit matrix, and $\chi_0^2$ is the
white noise intensity. In homogeneous states, the
Boltzmann--Fokker--Planck equation characterizing the evolution of a
 granular gas submitted to the stochastic external force $\mathbf{F}^{\mathrm{wn}}$ is
 written as \cite{VSK14}
\begin{equation}
\left(\partial_t -\frac{\chi_0^2}{2}\nabla_{\mathbf{v}}^2\right) f(\mathbf{v} , \bomega;t)={J[\mathbf{v}, \bomega|f(t)]}.
\label{BE}
\end{equation}
Above, $f(\mathbf{v},\bomega;t)$ is the velocity distribution
 function ($\mathbf{v}$ and $\bomega$ being the translational and angular velocities, respectively) and $J[\mathbf{v}, \bomega|f]$ is the collision integral in the
 (inelastic) Boltzmann equation for rough spheres, which accounts for
   the collision rules \cite{VSK14}
   \begin{equation}    \widehat{\bsigma}\cdot\mathbf{u}'=-\alpha\,\widehat{{\bsigma}}\cdot\mathbf{u},
\quad \widehat{{\bsigma}}\times\mathbf{u}'=-\beta \,
\widehat{{\bsigma}}\times\mathbf{u}.
     \label{collrule}
\end{equation}
Here, the primes denote postcollisional values,
$\widehat{{\bsigma}}$ is the unit collision vector joining the
centers of the two colliding spheres (from the center of particle 1 to
the center of particle 2) and
$\mathbf{u}=\mathbf{v}_1-\mathbf{v}_2-{\frac{\sigma}{2}}\widehat{{\bsigma}}\times({\bomega}_1+{\bomega}_2)$
is the relative velocity of the spheres at their contact point. The
coefficient of normal restitution $\alpha$ takes values between $0$
(completely inelastic collision) and $1$ (completely elastic
collision), while the coefficient of tangential restitution $\beta$
takes values between $-1$ (completely smooth collision, unchanged
angular velocities) and $1$ (completely rough
collision) \cite{B72}.

Given any one-particle function $A(\mathbf{v},\bomega)$, its average is defined
as $\langle A(t)\rangle=n^{-1}\int d \mathbf{v}\int d \bomega\,
A( \mathbf{v}, \bomega)f(\mathbf{v}, \bomega;t)$, where the number density is given by
$n=\int d \mathbf{v}\int d \bomega\, f( \mathbf{v}, \bomega;t)$.  The basic physical properties
are the translational ($T_t$), rotational ($T_r$), and total ($T$)
granular temperatures, i.e.,
\begin{equation}
\label{Tt,Tr}
{T_t={\frac{m}{3}}\langle v^2\rangle,\quad T_r={\frac{I}{3}}\langle
  \omega^2\rangle,} \quad
{ T=\frac{T_t+T_r}{2}=T_t\frac{1+\theta}{2}},
\end{equation}
where $I$ is the moment of inertia. We have introduced the temperature ratio $\theta\equiv T_r/ T_t$, which
is relevant for the analysis that follows and whose steady-state value
is independent of the driving amplitude $\chi_0^2$. The evolution
equations for $T_t$, $T_r$, and $T$ are
\label{Tidt}
\begin{equation}
\partial_t T_t-{m\chi_0^2}=-\xi_t T_t,\quad {\partial_t T_r=-\xi_r T_r},
\end{equation}
\begin{equation}
 \partial_t T-\frac{m\chi_0^2}{2}=-\zeta  T.
\end{equation}
The equations for $T_t$ and $T_r$ have been obtained by multiplying
both sides of Eq.\ (\ref{BE}) by the translational and rotational
kinetic energies, respectively, and integrating over all particle
velocity values. The parameters $\xi_t$ and $\xi_r$ are
\begin{equation}
\xi_t=-{\frac{m}{3n T_t}}\int d\mathbf{v}\int
d {\bomega}\,v^2{J[\mathbf{v}, {\bomega}| f]},
\end{equation}
\begin{equation}
\xi_r=-{\frac{I}{3n T_r}}\int d\mathbf{v}\int
d {\bomega}\,\omega^2{J[\mathbf{v},{\bomega}| f]},
\end{equation}
respectively. In general, neither $\xi_t$ nor $\xi_r$ does have a
definite sign, whereas the cooling rate,
\begin{equation} \zeta=\frac{\xi_t T_t+\xi_r
  T_r}{2T}=\frac{\xi_t+\xi_r{\theta}}{1+\theta},
  \end{equation}
is always positive
because energy is dissipated in collisions.

To proceed further, it is convenient to go to
dimensionless variables. Time is measured in a scale $\tau$,
\begin{equation}
  \tau=\frac{1}{2}\int_0^t dt'\,\nu(t'),\quad \nu(t)=4n\sigma^2 \sqrt{\pi
    T_t(t)/m},
    \end{equation}
which is roughly  the accumulated number of collisions per particle, because
$\nu(t)$ is the collision frequency. Dimensionless velocities are introduced as
\begin{equation}
  \mathbf{c}(t)\equiv\frac{\mathbf{v}}{\sqrt{2 T_t(t)/m}}, \quad
  \mathbf{w}(t)\equiv\frac{{\bomega}}{\sqrt{2 T_r(t)/I}},
\label{cw}
\end{equation}
a reduced  velocity distribution function as
\begin{equation}
\phi(\mathbf{c},\mathbf{w};\tau)\equiv \frac{1}{n}\left[\frac{4 T_t(t)
    T_r(t)}{m I}\right]^{3/2}f(\mathbf{v}, {\bomega};t),
\end{equation}
and the dimensionless collision kernel as
\begin{equation}
  \label{Cop}
\mathcal{J}[\mathbf{c},
\mathbf{w}|\phi(\tau)]=\frac{2}{n\nu(t)}\left[\frac{4 T_t(t) T_r(t)}{m I}\right]^{3/2} J[\mathbf{v}, {\bomega}|f(t)].
\end{equation}

In dimensionless variables, the evolution equations for the
temperatures can be written as Eqs.\ (\ref{evol_gamma}) and (\ref{evol_gamma2}) in the main
text. Therein, there appear the reduced collisional moments $\mu_{20}\equiv\mu_{20}^{(0)}$ and
$\mu_{02}\equiv\mu_{02}^{(0)}$, where
\begin{equation}
  \label{mupq}
\mu_{pq}^{(r)}(\tau)\equiv-\int d\mathbf{c}\int d\mathbf{w}\,
c^pw^q
(\mathbf{c}\cdot\mathbf{w})^r
{\mathcal{J}[\mathbf{c},\mathbf{w}|\phi(\tau)]}.
\end{equation}
Note that, aside from the nondimensionalizing factors, the production
rates $\xi_t$ and $\xi_r$ are basically identical to $\mu_{20}$ and
$\mu_{02}$, respectively. These are functionals of the whole
distribution function and thus the evolution equations for the
temperatures are not closed.

In order to close the dynamical equations, a formally exact expansion
in orthogonal polynomials can be performed \cite{VSK14}. For isotropic
states, we can expand the velocity distribution around the Maxwellian
$\phi_M(c,w) = \pi^{-3}e^{-c^2-w^2}$,
\begin{equation}\label{ap:phi_sonine}
\phi(\mathbf{c},\mathbf{w};\tau)=\phi_M(c,w)\sum_{j=0}^\infty\sum_{k=0}^\infty\sum_{\ell=0}^\infty a_{jk}^{(\ell)}(\tau)\Psi_{jk}^{(\ell)}(\mathbf{c},\mathbf{w}),
\end{equation}
where $\Psi_{jk}^{(\ell)}(\mathbf{c},\mathbf{w})$ are certain products
of Laguerre and Legendre polynomials. By normalization,
$a_{00}^{(0)}=1$, $a_{10}^{(0)}=a_{01}^{(0)}=0$, and the lowest
nontrivial coefficients are  those associated with moments of  degree four, namely

\begin{equation}\label{cumulant}
a_{20}^{(0)}=\frac{4}{15}\langle c^4\rangle-1,
\quad
a_{02}^{(0)}=\frac{4}{15}\langle w^4\rangle-1,
\end{equation}
\begin{equation}\label{a20}
 a_{11}^{(0)}=\frac{4}{9}\langle c^2w^2\rangle-1,\quad
a_{00}^{(1)}=\frac{8}{15}\left[\langle
  (\mathbf{c}\cdot\mathbf{w})^2\rangle-\frac{1}{3}\langle
  c^2w^2\rangle\right] ,
\label{b}
\end{equation}
which we call the fourth-order cumulants henceforth.

\textit{Maxwellian approximation.-} The simplest description is
obtained by substituting the Maxwellian velocity distribution into the
collision integrals (\ref{mupq}). Equivalently, one may consider that
all the nontrivial cumulant vanish in this approach, which yields
\begin{eqnarray} \label{mu2002M}
\mu_{20,M}&=1-\alpha^2+\frac{\kappa(1+\beta)}{(1+\kappa)^2}\left[2+\kappa(1-\beta)-\theta(1+\beta)\right], \label{mu20M}
\\
\mu_{02,M}&=\frac{\kappa(1+\beta)}{(1+\kappa)^2}\left[2+\kappa^{-1}(1-\beta)-\theta^{-1}(1+\beta)\right]. \label{mu02M}
\end{eqnarray}
where $\kappa\equiv 4I/m\sigma^2$ is the dimensionless moment of
inertia. Insertion of Eq.\ (\ref{mu2002M}) into the evolution
equations (\ref{evol_gamma}) and  (\ref{evol_gamma2}) in the main text gives rise to the
Maxwellian approximation.

\textit{First Sonine approximation.-} A more elaborate approximation
can be done by incorporating the lowest order cumulants, which we
defined in Eqs.\ (\ref{cumulant}) and (\ref{a20}), as the first corrections to
the Maxwellian.

A closed set of six coupled differential equations
can be obtained for $\theta(\tau)$, $\gamma(\tau)$,
$a_{20}^{(0)}(\tau)$, $a_{02}^{(0)}(\tau)$, $a_{11}^{(0)}(\tau)$, and
$a_{00}^{(1)}(\tau)$.
To do so, explicit---yet not exact---expressions
 for the collision integrals $\mu_{pq}^{(r)}$ with $p+q+2r=2$ and $4$
 are derived in terms of $\theta$ and those lowest order
 cumulants. These rather involved expressions can be found in the
 Supplemental Material of Ref.~\cite{VSK14}, and are thus omitted
 here. The resulting set of six differential equations can be
numerically solved with appropriate initial conditions for each
physical situation, as discussed in the main text. In this way, we
obtain the time evolution of the temperatures in the so-called first
Sonine approximation, to which we refer throughout this work.

\section{Computer simulations}
\label{app_B}
We use in this work data sets obtained
from computer simulations from two independent and different methods:
direct simulation Monte Carlo (DSMC) method, which obtains an exact
numerical solution of the relevant kinetic equation [in our case
Eq.\ (\ref{BE})] and molecular dynamics (MD) simulation, which solves
particles trajectories. A detailed description of the DSMC method may
be found elsewhere \cite{B94}. In our DSMC simulations, and in order to
reduce statistical noise in the temperature time evolution curves, we
have used an average of 100 statistical replicas of a system with
$2\times10^6$  particles. In the MD case, we have simulated
1000 inelastic hard spheres at a density $n\sigma^{3}=0.01$ and
averaged over 500 trajectories.

\newcommand{\newblock}{}

\bibliographystyle{unsrt}
\bibliography{KRWN}
\end{document}